\documentclass[twocolumn]{aastex631}
\usepackage{amsmath}
\usepackage{listings}
\usepackage{subfigure}
\usepackage{enumitem}
\let\tablenum\relax
\usepackage{siunitx}

\newcommand{\jpgilon}{\varepsilon}
\received{March 1, 2025}
\revised{June 17, 2025}
\accepted{\today}

\begin{document}

\title{Exoplanet Classification through Vision Transformers with Temporal Image Analysis}

\correspondingauthor{C. Swastik}
\email{swastik.chowbay@iiap.res.in}

\author{Anupma Choudhary}
\affiliation{Department of Mathematics and Computing, Indian Institute of Technology (Indian School of Mines), Dhanbad 826004, India}

\author{Sohith Bandari}
\affiliation{Department of Electronics Engineering, Indian Institute of Technology (Indian School of Mines), Dhanbad 826004, India}

\author[0000-0001-9569-2297]{B.S.Kushvah}
\affiliation{Department of Mathematics and Computing, Indian Institute of Technology (Indian School of Mines), Dhanbad 826004, India}

\author[0000-0003-1371-8890]{C. Swastik}
\affiliation{Dipartimento di Fisica, Universit\'a degli Studi di Milano, Via Celoria 16, 20133 Milano, Italy}
\affiliation{Indian Institute of Astrophysics, Koramangala 2nd Block, Bangalore 560034, India}
\affiliation{Pondicherry University, R.V. Nagar, Kalapet, 605014, Puducherry, India}

\begin{abstract}
{The classification of exoplanets has been a longstanding challenge in astronomy, requiring significant computational and observational resources. Traditional methods demand substantial effort, time, and cost, highlighting the need for advanced machine learning techniques to enhance classification efficiency. In this study, we propose a methodology that transforms raw light curve data from NASA's Kepler mission into Gramian Angular Fields (GAFs) and Recurrence Plots (RPs) using the Gramian Angular Difference Field and recurrence plot techniques. These transformed images serve as inputs to the Vision Transformer (ViT) model, leveraging its ability to capture intricate temporal dependencies. We assess the performance of the model through recall, precision, and F1 score metrics, using a 5-fold cross-validation approach to obtain a robust estimate of the model's performance and reduce evaluation bias. Our comparative analysis reveals that RPs outperform GAFs, with the ViT model achieving an 89.46$\%$ recall and an 85.09$\%$ precision rate, demonstrating its significant capability in accurately identifying exoplanetary transits. Despite using under-sampling techniques to address class imbalance, dataset size reduction remains a limitation. This study underscores the importance of further research into optimizing model architectures to enhance automation, performance, and generalization of the model.}
\end{abstract}

\keywords{Transit Method -- Exoplanet -- Convolution Neural Networks -- Recurrence Plots -- Vision Transformers}

\section{Introduction}
\label{Introduction}

The groundbreaking discovery of the first exoplanet orbiting a star similar to our Sun by \cite{1995IAUC.6251....1M} sparked a new era of research in the field of astronomy. Since then, thousands of exoplanets have been discovered, ranging in size and mass from those similar to Earth to massive, Jupiter-sized planets. With the detection of over 5800 exoplanets \citep{https://doi.org/10.26133/nea12}, scientists are now able to study the different populations of exoplanets and their host stars in greater detail \citep{2005ApJ...622.1102F,2007ARA&A..45..397U,2022AJ....164...60S,2022AJ....164..181U,2023AJ....165...34H,2025AJ....169...13P,2017AJ....154..108J}. By investigating the correlations between the properties of exoplanet host stars and the planets they host, researchers can gain a deeper understanding of how these planets formed and evolved. It's also worth noting that many exoplanet systems do not resemble our solar system, for instance, the prevalence of ``super-Earths" and ``mini-Neptunes" which are not found in our solar system, and ``hot Jupiters" which are extremely close to their host stars. This highlights that our solar system is somewhat atypical and that exoplanet formation and migration may have occurred differently when compared to our solar system.

Over time, various techniques have been developed for detecting exoplanets. The most direct way to detect exoplanets is direct imaging. The direct imaging technique uses high-contrast imaging and advanced technology like coronagraphs and adaptive optics, is typically used to find exoplanets in wider orbits ($\sim$ tens of AU), specifically young, self-luminous systems that are easier to detect \citep{2005A&A...438L..29C,2021AJ....161..114S,2023A&A...680A.114R,2024A&A...687A.257W,2025AJ....169...13P}. However, only a handful of planets ($\sim$70) have been detected by direct imaging. The majority of exoplanets detected so far are by transit, followed by radial velocity technique. The radial velocity detection technique, which also detected the first exoplanet, 51 peg, measures the tiny wobbles in a star's motion caused by the gravitational pull of an orbiting exoplanet, revealing the planet's presence and mass. The transit method, on the other hand detects exoplanets by observing the slight dimming of a star's light when a planet passes in front of it, revealing the planet's size and orbit \citep{borucki2010kepler}. They have been crucial in advancing our understanding of exoplanetary systems and are continually being refined to facilitate future discoveries \citep{wright2012exoplanet,perryman2018exoplanet,2023AJ....166...91S}. However, it's important to note that different methods have different biases and limitations. For example, the radial velocity (RV) and transit methods are currently the most successful methods for detecting exoplanets, but they are most effective for finding planets in close to moderate orbits (0.1-10 AU). On the other hand, the direct imaging technique is sensitive to find exoplanets in wider orbits ($\sim$ tens of AU), specifically young, self-luminous systems that are easier to detect \citep{2024AJ....167..270S}.

In a significant endeavor to expand our knowledge of exoplanets, several missions have been launched by various space agencies. The French Space Agency initiated the CoRoT mission in $2006$, pioneering space-based transit observations and discovering numerous exoplanets \citep{2006cosp...36.3749B}. Building on this success, the National Aeronautics and Space Administration (NASA) launched Kepler’s mission in $2009$, which played a vital role in expanding our knowledge of exoplanetary systems \citep{borucki2010kepler}. Kepler was established to monitor over $170,000$ stars in the Milky Way, especially in the Cygnus-Lyra area, with the main objective of finding Earth-sized exoplanets in the habitable zone of their host stars using the transit method \citep{borucki2016kepler}. By analyzing the slight dimming of starlight caused by planetary transits, Kepler substantially advanced the discovery and classification of exoplanets. This mission revealed a rich diversity in exoplanets' sizes, compositions, and orbital dynamics  \citep{borucki2010kepler,batalha2014exploring,borucki2016kepler}. Running until $2018$, Kepler significantly enhanced our understanding of exoplanetary systems, confirming over $2,600$ exoplanets and demonstrating the widespread presence of planets in our galaxy.

Despite the significant contributions of various space missions, the complexity and volume of Kepler data necessitated the development of advanced techniques for exoplanet classification. Early machine learning methods, such as random forests, were implemented to automate the classification of Kepler objects of interest, significantly reducing human effort and minimizing errors, thus leading to more efficient and accurate classification \citep{mccauliff2015automatic}. Furthermore, studies such as \cite{basak2021habitability} have shown that machine learning techniques can classify new exoplanets, but capturing complex patterns inside the large data set is still a challenging task. To address these limitations, \cite{lecun1998gradient} introduced Convolutional Neural Networks (CNNs), a subclass of deep neural networks specifically designed for image-based classification. CNNs consist of multiple layers of convolutional filters to autonomously learn and extract spatial hierarchies of features from input images. Each successive layer of filters is designed to extract complex features, from simple edges and textures to intricate patterns \citep{krizhevsky2012imagenet}. This hierarchical approach allows CNNs to effectively analyze and classify visual data with remarkable accuracy and efficiency \citep{szegedy2015going}.

Building on the limitations of CNNs in capturing long-term dependencies within data sets and the successful emergence of transformers in Natural Language Processing (NLP) inspired researchers, including  \cite{dosovitskiy2020image} to experiment with a standard Transformer by applying it directly to images, with the fewest possible modifications. As a result, the Vision Transformer (ViT) was developed, which offers a new approach to image-based classification as detailed by \cite{dosovitskiy2020image}. The fundamental concept of the model involves a sequence of visual patches as tokens for classifying the full image. It has achieved a benchmark performance in multiple image-based recognition tasks. In addition, transformer-based architectures have been successfully applied to numerous vision problems, including semantic segmentation \citep{9578646}, video understanding \citep{9710415}, image processing using ViT \citep{dosovitskiy2020image}, and object detection \citep{9709981}. Inspired by the rapid evolution of transformers, we have also implemented this in our methodology to examine its potential in exoplanet classification.

In the field of image-based classification,
\cite{eckmann1995recurrence} introduced Recurrence Plots (RPs) as a technique used to visualize the periodic nature inside the time series data. RPs convert time series data into a two-dimensional image where patterns and structures in the data can be more easily identified, facilitating the detection of hidden periodicities and the analysis of dynamic behavior over time \citep{eckmann1995recurrence}. In contrast, \cite{wang2015encoding} introduced the concept of Gramian Angular Fields (GAFs), which provide a method for transforming time series data into images by encoding angular information. GAFs represent time series data in a polar coordinate system, capturing the temporal correlations between different points. This transformation emphasizes the angular relationships between data points, which can be particularly useful for capturing cyclical patterns and trends \citep{wang2015encoding}.

Considering the potential of the above-mentioned transformation of 1D time series to 2D image representations, researchers have used them with various machine learning models in diverse fields like \cite{10178569,10290014} highlighted the potential of RPs and GAFs by projecting 1D Photoplethysmogram (PPG) signals into 2D images as input to ViT model for enhancing the overall performance and the results demonstrated a competitive prediction accuracy as compared to the current state-of-the-art. Subsequently, these advancements have enhanced the performance in tasks such as classification accuracy, anomaly detection, and forecasting \citep{10178569,10290014}. These techniques have also been applied in the field of astronomy, where they have been used for enhancing the efficiency of solar panels \citep{10320347}.

In this way, recent research studies have successfully integrated image representations with advanced modeling algorithms for time series data, showcasing their effectiveness in tasks related to image-based classification. Therefore, this study intends to build upon this foundation by integrating these representations as input in the ViT model to evaluate its potential in exoplanet classification. Further, we performed a comparative analysis between RPs and GAFs to assess their performance when integrated with the ViT model, providing a framework to explore further in the field of astronomy.

This paper is organized as follows. Section \ref{Introduction} offers an in-depth overview of the background of exoplanet research, detailing various missions and the machine learning techniques for exoplanet classification. In Section \ref{sec:DataPreprocessing}, we present our methodology, which encompasses the processes of data collection and preprocessing. Additionally, we provide details of the conversion of data into the preferred formats: RPs and  GAFs. This section provides a comprehensive overview of the techniques and strategies applied to prepare the data for subsequent analysis using ViT. Section \ref{Model Architecture} gives the subtle architectural details of ViT and hyperparameters used in the model. In Sections \ref{sec:ExoplanetClassification} and \ref{Results and Discussion}, we will provide detailed information on the data used in our study and present a comprehensive analysis of our comparative study of RPs and GAFs as inputs to our model. These sections elaborate on the techniques used for categorizing exoplanets, along with an in-depth discussion of the results obtained and their implications for the field of exoplanetary science. Section \ref{Conclusion} will culminate with a detailed discourse on the conclusions drawn from our comparative study, emphasizing that RPs have performed better on this dataset.


\section{Data Preprocessing}
\label{sec:DataPreprocessing}



\subsection{Light curves}
\label{Light curves}
NASA's Kepler mission, launched in 2009, was specifically designed to search for exoplanets by monitoring the brightness of over $156,000$ stars in a fixed field of view \citep{borucki2010kepler}. This mission deepens our understanding of exoplanetary systems by detecting subtle changes in stellar brightness caused by planetary transits over vast distances, typically known as light curves associated with a particular star, which could further be used to identify and classify extrasolar planets. Transiting dips are the reduction in flux values that provide valuable information about the transiting planet, whereas the width and depth of these dips provide insights about the transiting events. Specifically, the width of the transit provides information about the orbital period and distance from the star, while the depth of the dip gives a relationship between the radius of the star and the radius of the transiting planet. 
\begin{table}
	\centering
	\caption{Description of key attributes in the exoplanet dataset.}
	\begin{tabular}{lc}
		\hline
		\hline
		Attribute & Meaning \\
		\hline
		rowid & Identifier for each row in the TCE table \\
		kepid & Kepler ID \\
		tce\_plnt\_num & TCE planet number \\
		tce\_period & Orbital period \\
		tce\_time0bk & Transit epoch in BJD format \\
		tce\_duration & Transit duration \\
		av\_training\_set & Classification labels \\
		\hline
	\end{tabular}
	\label{tab:attributes}
\end{table}

The shape of the dip in a light curve reveals significant details about the transiting object and its orbital geometry. There are generally two types of dips- a U-shaped and a V-shaped. The U-shaped dip results from a central, edge-on transit, which occurs at an orbital inclination close to 90\si{\degree}, where the planet crosses near the center of the stellar disk. Conversely, a V-shaped dip signifies a grazing transit, where the planet partially crosses the stellar disk. While this also requires a high orbital inclination, it is slightly less edge-on compared to a central transit. Additionally, V-shaped dips are characteristic of eclipsing binary stars, where two stars pass in front of each other \citep{magliano2023tess}.


In this study, we used the Kepler light curve data for the exoplanet classification, consisting of labeled Threshold Crossing Events (TCEs). A threshold-crossing event occurs when the Kepler pipeline \citep{jenkins2010overview} detects periodic dips in stellar brightness, indicating a potential planetary transit. By analyzing these dips (TCEs), we can distinguish between actual exoplanet transit signals and other astrophysical phenomena or instrumental noise. 

\begin{figure}
    \centering
    \includegraphics[width=1\linewidth]{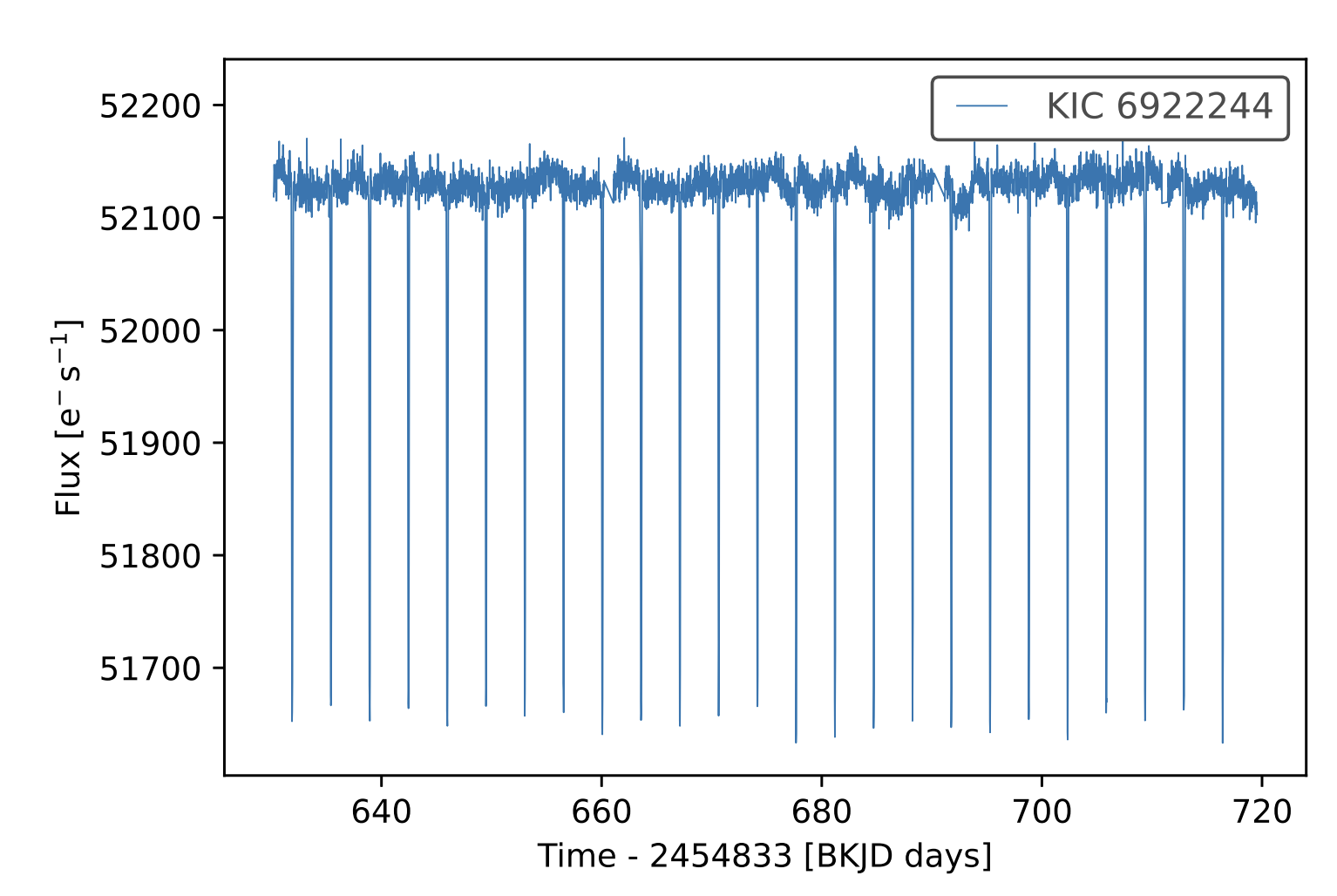}
    \caption{This diagram illustrates the transit crossing events observed in the light curve data for (KIC) candidate with ID 6922244. The plotted
points represent the variations in flux as the exoplanet transits its host
star, causing periodic dips in brightness.}
    \label{KIC_6922244}
\end{figure}

The TCEs shown in Figure \ref{KIC_6922244} illustrate notable reductions in the normalized flux for the Kepler ID \lq \texttt{KIC 6922244}\rq\ over time. Here, \texttt{KIC} refers to the Kepler Input Catalog. The sequence of periodic dips (TCE) or straight line patterns, shown in Figure \ref{KIC_6922244}, indicate the possibility of actual exoplanetary signals. This gives insight into the role of TCEs in classifying potential exoplanet candidates. We will use these labeled TCEs for the classification. The labels, assigned through human vetting, provide a reliable basis for distinguishing between true exoplanet candidates and false positives.

In this analysis, we accessed labeled TCEs data from the \cite{https://doi.org/10.26133/nea12}, specifically the Autovetter Planet Candidate Catelog for Q1-Q17 DR24 \citep{catanzarite2015autovetter}. The data encompasses 20,367 TCEs, each with a set of attributes or columns, and we selected specific columns relevant to our analysis. Table \ref{tab:attributes} represents the seven key attributes with their usual meanings.

\begin{table}
	\centering
	\caption{Configuration of the dataset used in the study, detailing the number of training examples for each label.}
	\begin{tabular}{lc}
		\hline
		\hline
		\multicolumn{2}{c}{Data Configuration} \\
		\hline
		Labels & Number of Training Set Examples \\
		\hline
		AFP & 9596 \\
		NTP & 2541 \\
		PC & 3600 \\
		UNK & 4630 \\
		\hline
	\end{tabular}
	\label{tab:data_configuration}
\end{table}

\begin{figure*}[ht!]
    \centering
    \gridline{
        \fig{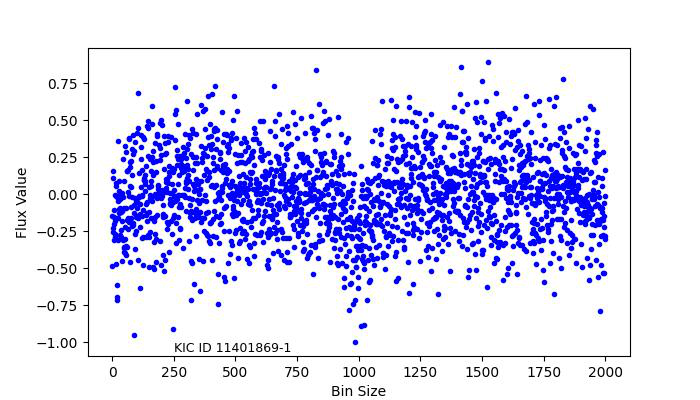}{0.45\textwidth}{(a)}
        \fig{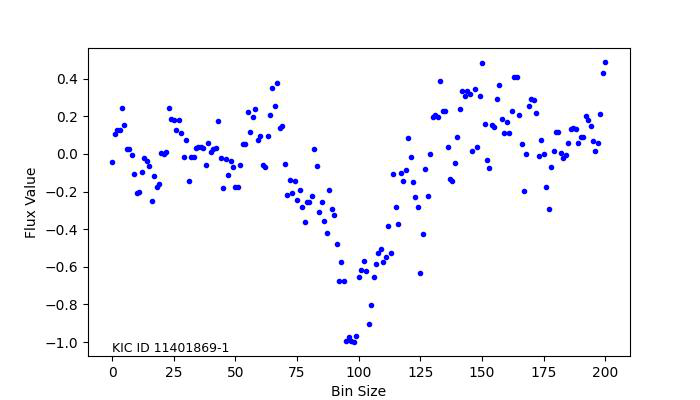}{0.45\textwidth}{(b)}
    }
    \gridline{
        \fig{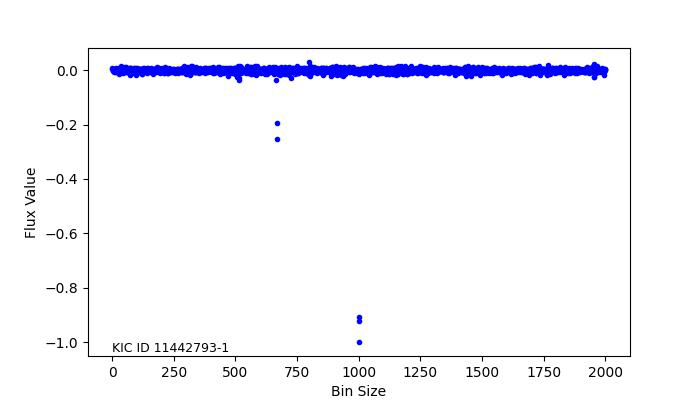}{0.45\textwidth}{(c)}
        \fig{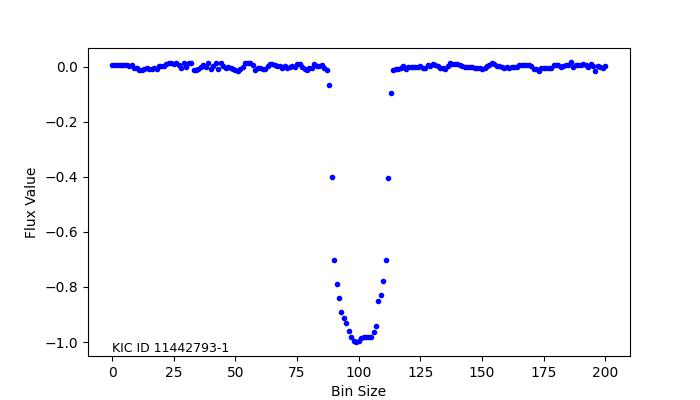}{0.45\textwidth}{(d)}
    }
    \caption{(a) and (c) represent the global light curves of AFP and PC candidates, respectively, consisting of full-length data with a resolution of $2001$ bin size. (b) and (d) represent the local light curves for the same candidates, which are fixed-length representations focusing on the transit region with a bin size of $201$. Each plot highlights the key periodic features distinguishing these astrophysical phenomena.}
    \label{fig:global_and_local}
\end{figure*}

The most crucial attribute, $av\_training\_set$, consists of four distinct labels: non-transiting phenomena (NTP), astrophysical false positives (AFP), planet candidates (PC), and unknown (UNK). NTP includes phenomena that are not related to transits, such as stellar variability, while AFP signifies signals initially thought to be transits but later identified as noise or other astrophysical events. PC denotes probable transiting exoplanets, and UNK represents signals that cannot be clearly classified. These labels provide a comprehensive framework for categorizing various astrophysical signals. The detailed configuration is given in Table~\ref{tab:data_configuration}. For binary classification purpose, the dataset is further categorized into \lq\lq planet\rq\rq\ candidates and \lq\lq not-planet\rq\rq\  candidates, the latter comprising NTP and AFP. We ignored TCEs with the \lq\lq unknown\rq\rq\ label (UNK). Table \ref{tab:data_configuration} indicates a substantial class imbalance, with  3,600 \lq\lq planet\rq\rq\  candidates and the remainder classified as \lq\lq not-planet\rq\rq\ (NTP/AFP) candidates. \cite{catanzarite2015autovetter} reported that the inclusion criteria for these labels were developed by a precise procedure that included manual vetting and diagnostic evaluations. In this study, we assume these labels as the ground truth, considering that the potential impact of any inaccuracies on our model is negligible.

The light curves of the corresponding stars were downloaded through the \cite{https://doi.org/10.17909/T9059R}, with each light curve generated through the Kepler pipeline containing integrated flux measurements at $29.4$-minute intervals over up to four years.

Following the procedure of \cite{shallue2018identifying}, we performed additional stjpg on generated light curves. First, we removed data points corresponding to transits of any confirmed planets within the system. Subsequently, we flattened the light curves by fitting a basis spline to the data and normalized them by dividing each light curve by the best-fit spline. Finally, the flattened light curves were phase-folded on the TCE period and binned to produce a 1D vector. These binned and phase-folded light curves corresponding to each TCE provide local and global curves.

Figure \ref{fig:global_and_local} presents the global and local curves generated from the original light curve. The global light curve captured broad trends and changes throughout the event based on the photometric measurements, while the local light curve was concentrated to capture changes around the transit dips. Figure \ref{fig:global_and_local} provides the visual representation of both curves to evaluate the effectiveness of each curve in capturing critical patterns within the data. Figure \ref{fig:global_and_local}(c) presents the global curve of the KIC ID 11442793-1, in which identifying narrow transit was a difficult task or may provide us with a wrong prediction, whereas the local light curve (Figure \ref{fig:global_and_local}(d)) makes it easier by providing subtle signals inside the data. Therefore, considering its effectiveness in capturing subtle patterns, we have used the local light curve as the primary approach in this study, which makes it different from the one that incorporates local and global views, such as \cite{shallue2018identifying}.

Further, we transformed the local curve data set into RPs and GAFs representations to feed them for training the model and evaluating its performance using evaluation metrics to compare the effectiveness of each in identifying hidden patterns within the data.

\subsection{Recurrence Plots (RPs)}

RPs analysis is a technique deduced from nonlinear dynamics and is particularly appropriate to complex signals \citep{6729553}. It provides a strong alternative for demonstrating and studying the periodic signals. RPs are a visual matrix showing the point at which a phase space trajectory returns to a state already visited. This method provides insight into viewing and identifying hidden repetitive patterns that exhibit the characteristics of orbital movements of exoplanet data, which enables the analysis of $m$-dimensional trajectories within a two-dimensional phase space. We define a recurrence matrix R, where each member is used to construct an RPs \citep{6729553}. 

\(R_{k,l}\) is defined by the proximity of state vectors 
$x(k)$ and $x(l)$ by the following formula:  

\begin{equation}
	R_{k,l} = \phi(\jpgilon - \|x(k) - x(l)\|), \quad x(\cdot) \in \mathbb{R}^m, \quad k, l = 1, \ldots, N
\end{equation}

The state vectors at indices $k$ and $l$ are represented by $x(k)$ and $x(l)$ respectively, and the symbol $\|.\|$ specifies the application of norm, between these observations. The closeness threshold is set by the parameter $\jpgilon$, and the Heaviside function, that is represented by $\phi$ in Equation 2, is as follows:
\begin{equation}
	\phi(z) =  \begin{cases} 0, & \text{if } z < 0 \\ 1, & \text{otherwise} \end{cases}
\end{equation}

\begin{figure}
	\centering
	\includegraphics[width=0.45\linewidth]{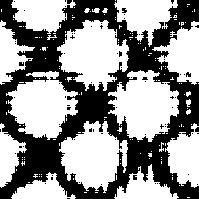}
	\includegraphics[width=0.45\linewidth]{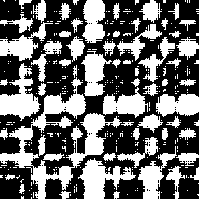}
	\includegraphics[width=0.45\linewidth]{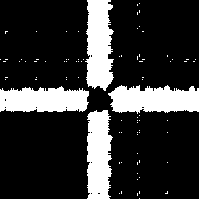}
	\includegraphics[width=0.45\linewidth]{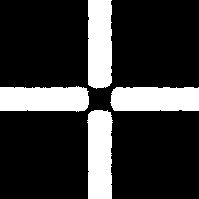}
	\caption{RPs for Exoplanet Classification. The upper two frames represent RPs for \lq\lq not-planet\rq\rq\ candidate cases, showing irregular patterns indicative of false positives. The lower two frames depict RPs for \lq\lq planet\rq\rq\ candidates, displaying clear, periodic structures consistent with true exoplanets.}
	\label{fig:RP}
\end{figure}

Equation 1 states that if the $m$-dimensional trajectory of the time series at time $k$ lies close to time $l$, then value 1 is assigned to \(R_{k,l}\); otherwise, 0 is assigned. Graphically, it is represented by an N $\times$ N matrix, with pixels corresponding to 1 being colored as black, showing recurrence behavior, and those relating to 0 being white. For generating RPs, we choose a threshold of 0.65, a time delay of 2, and a value of $m=2$, and apply the Euclidean distance to determine recurrences between the embedded state vectors, as it captures more temporal dynamics since each point is now represented as a 2D state vector. The produced RPs using these parameters have dimensions of 199 $\times$ 199 pixels, as shown in Figure \ref{fig:RP}. This visualization offers a powerful tool for identifying periodic patterns and randomness within time series data, with varied degrees of structure and complexity evident across different systems.
For example, the two lower frames in Figure \ref{fig:RP} indicate real exoplanet prospects with regular transit properties, indicative of stable exoplanetary systems. Conversely, systems characterized by high disturbance or high stellar activity may be represented by more disorganized and fractured plots (refer to the upper two frames of Figure \ref{fig:RP}), which could lead to false positive detections of exoplanets.

These RPs were generated using some essential \texttt{Python} libraries such as NumPy, PyRQA and matplotlib, where Recurrence Quantification Analysis (RQA) offers several measures to quantify the structures within the RPs, such as determinism, recurrence rate, and entropy \citep{rawald2017pyrqa}. These metrics can elucidate the nature of the dynamical systems observed, providing insights into the stability and predictability of exoplanetary orbits.  Due to its outstanding suitability for handling non-stationary and non-linear data, RQA is particularly well-suited for analyzing light curves, where noise and observational gaps frequently provide substantial obstacles.

\subsection{Gramian Angular Fields (GAFs)}
\label{Gramian Angular Fields}
\cite{wang2015encoding} proposed GAFs to convert time series data to visual patterns. GAFs are effective in capturing long-term dependencies, sparsity, and subtle signals, so we have used the GAFs method for converting the 1D time series to 2D visual representations to provide a framework for images wherein the image is characterized by angular coordinates rather than Cartesian coordinates. The encoding process first normalizes the time series data into the range of [-1,1]. Normalized input is transformed into polar coordinates to retain the temporal information of the input. After this, the trigonometric cosine function is used for temporal correlation by comparing each point in polar coordinate with every other point to generate a Gramian matrix of dimension n $\times$ n, where n is the length of time series data.\\
\begin{figure}
	\centering
	\includegraphics[width=0.45\linewidth]{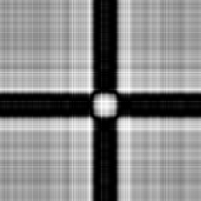}
	\includegraphics[width=0.45\linewidth]{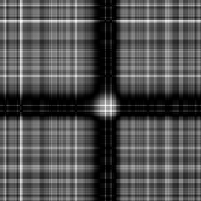}
	\includegraphics[width=0.45\linewidth]{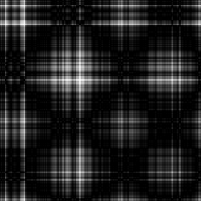}
	\includegraphics[width=0.45\linewidth]{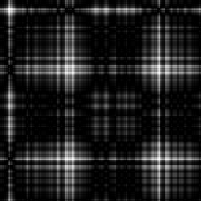}
	\caption{GAFs for Exoplanet Classification. The upper two frames represent GAFs for \lq\lq planet\rq\rq\ cases, characterized by distinct periodic patterns indicative of planetary transits. The lower two frames show GAFs for \lq\lq not-planet\rq\rq\ candidate cases, which lack the clear periodic structures seen in true exoplanet detections, indicating the absence of exoplanets.}
	\label{fig:GAFs}
\end{figure} 
Given a time series \( Y = \{y_1, y_2, \ldots, y_n\} \) of \( n \) real-valued observations, we first normalize the values into the range \([-1, 1]\) using the following equation:

\begin{equation}
	\tilde{y_i} = \frac{2y_i - \{\max(Y) + \min(Y)\}}{\max(Y) - \min(Y)}
\end{equation}

After normalization, each value in the time series is encoded into a polar coordinate system. Specifically, each value \(\tilde{y_i}\) is mapped to an angular cosine value, representing the angle \(\theta_i = \arccos(\tilde{y_i})\), while the temporal index is mapped to a radial coordinate. This mapping establishes a unique angular position for each time point in the series.

We then computed the Gramian Angular Field, where each element of the matrix \( G \) is defined as the trigonometric sum of angles between time stjpg:

\begin{equation}
G_{i,j} = \cos(\theta_i + \theta_j)
\end{equation}

The resulting symmetric GAFs matrices encapsulate the temporal patterns and underlying relationships observed in the exoplanetary time series data. Following the aforementioned methodological approach, a set of images with the resolution of 201 $\times$ 201 was generated, which is depicted in Figure \ref{fig:GAFs}, where the top two plots illustrate the periodic behavior of planet candidates by capturing subtle signals and the bottom two plots represent cases where no transiting planet candidates were detected. 

\section{Model Architecture}
\label{Model Architecture}
The classification of exoplanets using Kepler data is inherently challenging due to the intricate patterns embedded within light curves, which often contain deviations corresponding to planetary transits. Traditional approaches, such as CNNs, have demonstrated significant efficacy in various image-based classification tasks. However, CNNs use localized convolution operations, it may be difficult for them to capture the global patterns and long-term dependencies due to their limited scope.

To address these limitations and leverage its effectiveness in various image-based tasks, we have implemented ViT \citep{dosovitskiy2020image} in the exoplanet classification. ViT can be an effective approach to image-based classification by using a transformer architecture that treats an image as a sequence of patches, similar to the way words are processed in a sentence. Each patch is embedded into a token and processed through multiple layers of self-attention mechanisms, enabling ViT to capture both local and global relationships within the data. We have used RPs and GAFs as classification input tokens, allowing ViT to analyze Kepler data and classify exoplanets effectively.

Its architecture begins by dividing an input image $I$ into non-overlapping square patches; our model typically uses size $8 \times 8$. Each patch is flattened into a vector representing the pixel data, which is then projected into a higher-dimensional space via a learnable linear transformation matrix, $\mathbf{E}$. These patch embeddings are analogous to word embeddings in NLP. Alongside these tokens, a special classification token $[cls]$ is introduced at the beginning of the sequence. This token will later used to accumulate and represent the overall information of the image, which will be used for the final classification output. Moreover, positional encodings are added to each embedding to ensure the spatial relationships between patches are maintained, providing critical context to the self-attention mechanism \citep{dosovitskiy2020image}.

\begin{figure}[t]
	\centering
	\includegraphics[width=1\columnwidth]{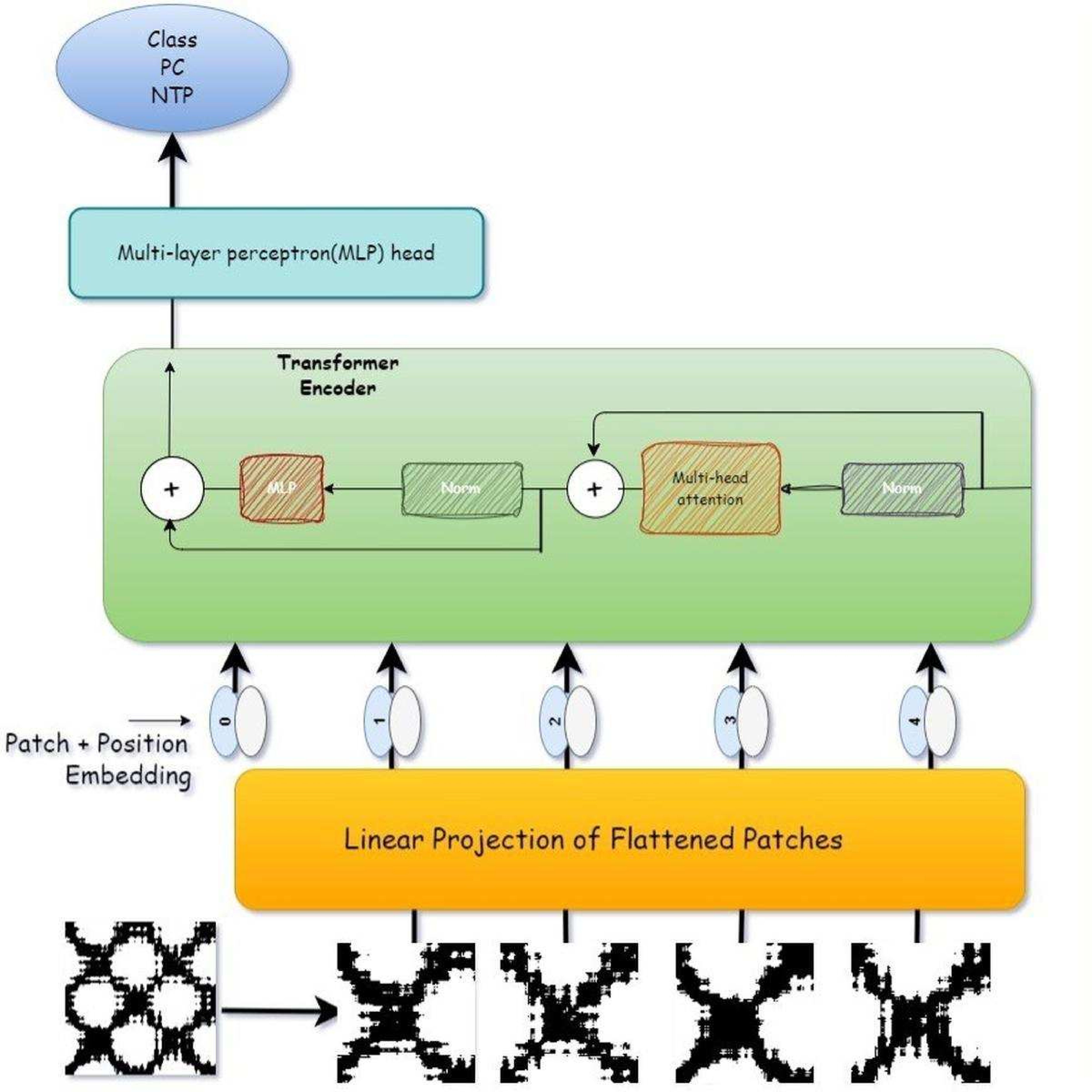}
	\caption{Architecture of the Vision Transformer (ViT) Model.}
	\label{Vision Transformer Model }
\end{figure}

Once the input sequence is ready, it is fed into the transformer encoder, a structure derived from the architecture proposed by \cite{vaswani2017attention, chi2020relationnet++}. In our model, we used ten such transformer layers. Each encoder layer has two primary components: a multi-head self-attention mechanism and a feed-forward network (FFN), as shown in Figure \ref{Vision Transformer Model }. Using four heads per layer, this mechanism enables the model to capture complex interdependencies and global features from different regions of image, essential for identifying subtle planetary signals. The multi-head self-attention mechanism allows every token to attend every other token in the sequence, capturing long-range dependencies and relationships between patches. It computes attention scores using the query, key, and value matrices obtained from the input embeddings. In the model architecture, the input embeddings of dimension 32 are first projected to a higher-dimensional space of dimension 128 to calculate the query, key, and value matrices across four attention heads. The concatenated output from all heads is then projected back to the original dimension 32, resulting in a higher parameter count for the multi-head attention layer, as shown in Table \ref{appendix:Model Architechture} in the Appendix. By performing multiple self-attention operations in parallel, the model can understand intricate relationships across different parts of the image.

After the self-attention stage, the feed-forward network, which consists of two fully connected layers (64 and 32 neurons) with a Gaussian Error Linear Unit (GELU) activation function, further refines the information extracted by the attention mechanism. The GELU is a non-linear activation function that combines the properties of the ReLU (Rectified Linear Unit) and Gaussian distributions. Mathematically, it is defined as $\text{GELU}(x) = x \cdot \Phi(x)$, where $ \Phi(x)  $ is the cumulative distribution function of a Gaussian distribution. This function provides a smoother transition by gradually adjusting the activation, unlike ReLU, which immediately sets all negative values to zero. The GELU function models data more accurately by weighing the input data's transformation based on its estimated probability distribution, allowing it to smoothly activate neurons rather than switching them off completely. This characteristic makes GELU particularly effective in deep learning architectures like ViT, where the diversity and complexity of the input data require nuanced processing. GELU helps in smoothing the learning pathway, allowing for better handling of the non-linearities found in the data, thus contributing to more robust and higher-quality feature extraction.

At the end of the encoder sequence, the classification token, which now contains aggregated global information from all patches, passes through a multi-layer perceptron (MLP) head. This MLP head, consisting of layers with $2048$ and $1024$ units, uses the refined global representations to determine the final classification. Each sub-layer in the encoder, both the self-attention and the feed-forward layers, includes residual connections and layer normalization, which help stabilize the training process by improving gradient flow.

In this study, the ViT model was initialized with random weights and trained using the AdamW optimizer with hyperparameters \( \beta_1 = 0.9 \), \( \beta_2 = 0.999 \), a learning rate of \( 0.0001 \), and a weight decay of \( 0.00001 \) to ensure model's optimization and regularization. Input images were resized to \( 192 \times 192 \) and divided into non-overlapping patches of size \( 8 \times 8 \), resulting in 576 patches per image. Each patch was flattened and linearly projected into an embedding space with a dimensionality of \( 32 \). 



\section{Exoplanet Classification}
\label{sec:ExoplanetClassification}
A structured implementation of the above methodology was performed to handle complex astronomical data for exoplanet classification. To address this challenge, we conducted a comparative analysis of the ViT model based on primary inputs: RPs and GAFs. This study investigates how using different input types enhances the reliability and robustness of exoplanet classification with the model.

To train our model for exoplanet classification, we adopted a systematic approach inspired by \cite{shallue2018identifying}. The dataset was randomly partitioned into three distinct subsets to ensure a comprehensive and unbiased evaluation of the model. The partitioning was as follows:

\begin{itemize}
	\item \textbf{Training Set}: Comprising 80\% of the total dataset, this subset was used to train the model. In the case of RPs as input, the dataset comprised $2,848$ \lq\lq planet\rq\rq\ candidates and $9,741$ \lq\lq not-planet\rq\rq\ candidates. Conversely, when using GAFs as input, the dataset consisted of $2,869$ \lq\lq planet\rq\rq\ candidates and $9,720$ \lq\lq not-planet\rq\rq\ candidates. The substantial size of this set ensured that the model had sufficient data to learn and generalize from a wide range of examples.
	\vspace{0.3cm}  
	\item \textbf{Validation Set}: This subset, comprising 10\% of the dataset, was used for hyperparameter tuning and model validation. In the case of RPs as input, the dataset comprised $382$ \lq\lq planet\rq\rq\ candidates and $1,192$ \lq\lq not-planet\rq\rq\ candidates. while for GAFs, it comperised $359$ \lq\lq planet\rq\rq\ candidates and $1,215$ \lq\lq not-planet\rq\rq\ candidates. The validation set played a crucial role in preventing overfitting by providing a basis for tuning the model's parameters and assessing its performance during training.
	\vspace{0.3cm}  
	\item \textbf{Test Set}: The remaining 10\% of the dataset was reserved for testing the model's performance. For RPs, this subset included 370 \lq\lq planet\rq\rq\ candidates and 1,204 \lq\lq not-planet\rq\rq\ candidates. For GAFs, the test set consisted of 372 \lq\lq planet\rq\rq\ candidates and 1,202 \lq\lq not-planet\rq\rq\ candidates. As detailed by \cite{shallue2018identifying}, the test set provides an unbiased evaluation of the model's accuracy and generalization capabilities.
\end{itemize}

The data for RPs and GAFs was partitioned to ensure that each split--training, validation, and testing--contained the same instances across both representations; however, slight differences in the number of planet and not-planet candidates between the two input types were due to the independent generation of RPs and GAFs images before partitioning into training, validation and testing dataset. Additionally, the models were trained separately on each data representation. To ensure the reliability of results, we have performed cross-validation. This ensures that the performance of the model with both inputs is comparable. Given the significant class imbalance in the dataset, evaluating the model's performance based solely on accuracy would be inappropriate. Therefore, we assessed the model's performance using recall and precision metrics. These metrics provide a more accurate and meaningful evaluation of the model's ability to handle an imbalanced data set. Precision and recall are critical metrics; Precision measures the model's ability to correctly identify relevant instances (true positives), while recall assesses the model's effectiveness in capturing all relevant instances. The F1-score, as the harmonic mean of Precision and recall, provides a balanced metric that addresses the trade-offs between these two aspects. The metrics are defined as:

Recall (Sensitivity or True Positive Rate):
\begin{align}
	\text{Recall} &= \frac{\text{TP}}{\text{TP} + \text{FN}}
\end{align}

Precision (Positive Predictive Value):
\begin{align}
	\text{Precision} &= \frac{\text{TP}}{\text{TP} + \text{FP}}
\end{align}

In these formulas:
\begin{itemize}
	\item TP (True Positives) refers to the number of correctly identified planet candidates.
	\item FN (False Negatives) refers to the number of actual planet candidates that were incorrectly classified as not-planet candidates.
	\item FP (False Positives) refers to the number of not-planet candidates that were incorrectly classified as planet candidates.
\end{itemize}

Upon evaluation of the test dataset, the model with RPs as input exhibited significant performance as compared to GAFs by achieving a recall of 89.46\% and a precision of 85.09\%. These results underscore the model's effectiveness when using RPs as input, demonstrating its ability to identify planet candidates accurately. This validates the model's utility for classification tasks within the context of the Kepler-DR $24$ dataset.


\section{Results and Discussion}
\label{Results and Discussion}
We present a comprehensive analysis of ViT performance in time-series data, which not only identifies the strengths of the proposed methodology but also points out its shortcomings, paving the way for further research and improvements in these areas. 

 In this study, we conducted a comparative analysis between two types of inputs, RPs, and GAFs, by using a ViT model for the classification of exoplanets from time-series data. The performance of the model was evaluated over $100$ epochs, and it took approximately 3 to 4 hours to train the model, which consists of $\sim$ 2.39 million parameters.  This computationally intensive task was executed using an NVIDIA DGX-1 8X Tesla V100 GPU, equipped with 32 GB of memory, and supports Tensor Cores, enhancing the performance of machine learning-related tasks. The following sections provide a detailed analysis and interpretation of the results, supported by visual aids.
\subsection{Performance Analysis of RPs}

\begin{table}
	\caption{General model performance for RPs and GAFs, evaluated using Precision, Recall, and F1-score.}
	\label{tab:merged_metrics_1}
	\centering
	\begin{tabular}{lcc}
		\hline\hline
		Metric & RPs & GAFs \\
		\hline
		Precision & 0.8509 & 0.7513 \\
		Recall & 0.8946 & 0.7956 \\
		F1-score & 0.8722 & 0.7728 \\
		\hline
	\end{tabular}
\end{table}

The metrics for evaluating the model performance with RPs as input are summarized in Table \ref{tab:merged_metrics_1}. The recall, precision, and F1-score values are $0.8946$, $0.8509$, and $0.8722$, respectively. These values indicate a balanced trade-off between precision and recall, highlighting the model's proficiency in minimizing false positives while effectively identifying true positives.
\begin{figure}[h]
	\centering
	\includegraphics[width=1\columnwidth]{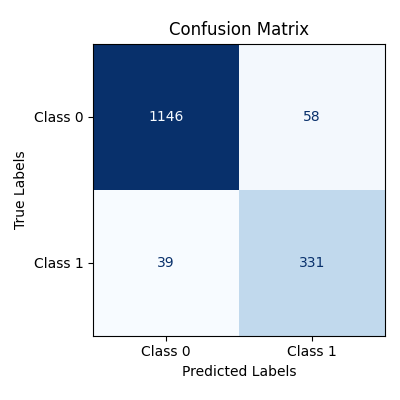}
	\caption{Confusion Matrix for RPs dataset, illustrating model performance. Class $0$ represents \lq\lq not-planet\rq\rq\ candidates, while Class $1$ represents \lq\lq planet\rq\rq\  candidates.}
	\label{fig:confusion_matrix_rp}
\end{figure}

Further supporting these results, the confusion matrix for the model shown in Figure \ref{fig:confusion_matrix_rp} visually illustrates its classification capabilities. The matrix reveals that the model accurately identified a significant number of both \lq\lq not-planet\rq\rq\  and \lq\lq planet\rq\rq\  candidates, with relatively few misclassifications. This underscores the model's effectiveness in distinguishing between the classes.

To validate the model’s robustness, we applied cross-validation, a widely used technique that assesses the model performance on independent datasets. Specifically, we used 5-fold cross-validation, where the dataset is split into five equal parts (folds). In each iteration, four folds are used for training, and one fold is used for testing iteratively, ensuring that every fold is used as a test set once. Cross-validation is crucial because it provides a more reliable estimate of model performance by reducing the variance associated with a single split of the data.

The cross-validation results yielded an average recall of 0.8810 and an average precision of 0.8426. To validate the consistency of the cross-validation results across all folds, we calculated the standard deviation and variance for each metric, as presented in Table \ref{tab:rp_stats}. The low value of standard deviation indicates minimal variability across all folds. Similarly, the small variance highlights the robustness irrespective of the specific fold partitioning. These results are consistent with the initial metrics, affirming the model's stability and reliability across different data subsets. By using cross-validation, we ensure that the model's performance is not overly dependent on a particular subset of the data, thereby improving its generalizability to unseen datasets.

However, while cross-validation confirms the model's robustness, it also brings attention to the persistent challenge of class imbalance in the data set. In this context, the number of \lq\lq not-planet\rq\rq candidates substantially exceeds that of \lq\lq planet\rq\rq candidates. This imbalance can lead to biased model predictions, where the model is inclined to the majority class, thereby reducing its sensitivity to detecting minority class instances. To mitigate this issue, an under-sampling technique was applied to the original dataset by randomly reducing the number of instances in the majority class to achieve a balanced dataset. This approach is designed to enhance the model’s capacity to accurately identify the minority class, thereby improving overall classification performance.

Following the application of under-sampling, the model was trained again and the performance was re-evaluated following the training process outlined in Section ~\ref{sec:ExoplanetClassification}. As shown in Table ~\ref{tab:merged_metrics}, there was a slight decrease in precision to 0.8426, recall to 0.8756, and the F1-score remained stable at 0.8587, indicating that the balance between precision and recall was effectively maintained.

\begin{table}[h]
\caption{Statistical Measures for 5-fold cross-validation (RPs).}
\label{tab:rp_stats}
\centering
\begin{tabular}{lccc}
\hline\hline
Metric    & Mean  & Standard Deviation & Variance \\ \hline
Precision       & 0.8426       & 0.0153                  &  0.0002         \\ 
Recall          & 0.8810        & 0.0149                   &  0.0002          \\
F1 Score        & 0.8614       &  0.0150                  & 0.0002         \\\hline
\end{tabular}

\end{table}

\begin{table}[h]
\caption{Statistical Measures for 5-fold cross-validation (GAFs).}
\label{tab:gaf_stats}
\centering
\begin{tabular}{lccc}
\hline\hline
Metric    & Mean  & Standard Deviation & Variance \\ \hline
Precision      & 0.7756        & 0.0470                   & 0.0022         \\ 
Recall         & 0.8102        & 0.0468                   & 0.0021        \\
F1 Score       & 0.7925        & 0.0468                   & 0.0021         \\\hline
\end{tabular}
\end{table}

\begin{figure}
	\centering
	\includegraphics[width=1\columnwidth]{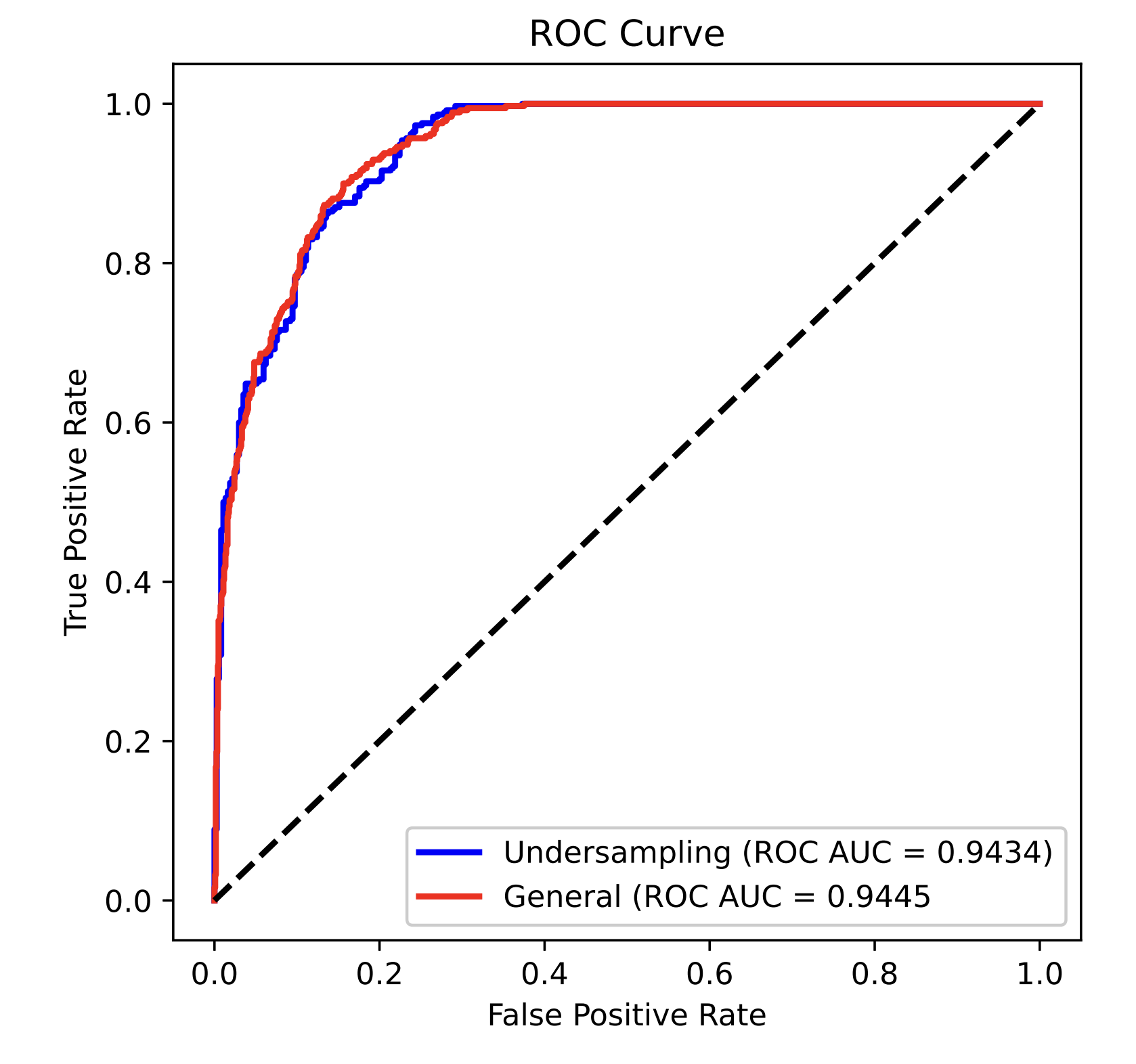}
    \caption{ROC Curve for the RPs-based model, comparing the performance of the general model (red) and the under-sampled model (blue).}
    \label{fig:roc_curve_rp}
\end{figure}

Additionally, the Receiver Operating Characteristic (ROC) curve offers a graphical representation of the model's diagnostic ability by plotting the True Positive Rate (TPR) on the y-axis and the False Positive Rate (FPR) on the x-axis, allowing for an intuitive assessment of the model's ability to distinguish between the positive and negative classes. A model that perfectly separates the classes would have a ROC curve that reaches the top-left corner of the plot, indicating a high TPR and a low FPR. As illustrated in Figure \ref{fig:roc_curve_rp}, the ROC curve showcases the model’s performance, with the diagonal line representing a random classifier, where TPR equals FPR and indicates no discriminatory power. The ROC curve for the general model reaches an ROC value of $0.9445$, significantly exceeding the diagonal line and reflecting strong predictive power. Although the ROC value for the under-sampled model slightly decreased to $0.9434$. The model demonstrates robustness in handling class imbalance, showing strong classification performance with RPs as input even without using under-sampling techniques.

\subsection{Performance Analysis of GAFs}

The model's performance using GAFs as input is also summarized in Table \ref{tab:merged_metrics_1}. The model achieved a precision of $0.7513$, a recall of $0.7956$, and an F1 score of $0.7728$. Although these values are slightly lower than those obtained with RPs, still reflect strong classification capabilities, demonstrating the model's effectiveness in identifying and capturing relevant instances within the dataset. 
\begin{figure}
	\centering
	\includegraphics[width=1\columnwidth]{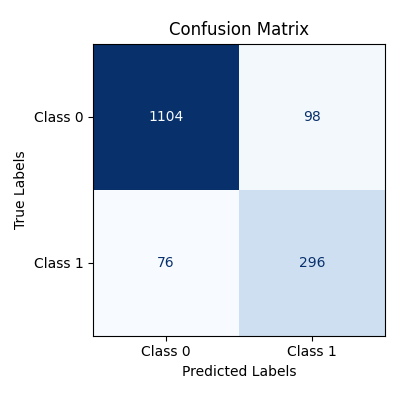}
	\caption{Confusion Matrix for GAFs dataset, illustrating general model performance. Class $0$ represents \lq\lq not-planet\rq\rq\ candidates, while Class $1$ represents \lq\lq planet\rq\rq\  candidates.}
	\label{fig:confusion_matrix_gaf}
\end{figure}

Further validation of these results is provided by the confusion matrix shown in Figure \ref{fig:confusion_matrix_gaf}. The model correctly identified a substantial number of not-planet candidates (1104) and planet candidates (296), with relatively few misclassifications (98 false positives and 76 false negatives). This distribution supports the model's significant classification accuracy and effectiveness in distinguishing between positive and negative classes, corroborating the values of the metrics presented in Table \ref{tab:merged_metrics_1}.

To ensure the robustness of the model, we performed a cross-validation technique. The cross-validation results for the model with GAFs as input produced an average recall of $0.8102$ and an average precision of $0.7756$. Following the analysis performed for RPs, we also calculated the same statistical measures for GAFs as presented in Table \ref{tab:gaf_stats} to thoroughly assess the consistency and reliability of its cross-validation results. The low standard deviations and variances confirm that the performance metrics for GAFs also exhibit minimal variability across folds. Although slightly lower than the general results, the values demonstrate the consistent performance of the model and confirm its robustness in handling various data subsets. This consistency across different data splits underscores the reliability and stability of the model when using GAFs as input.

To address the class imbalance in dataset, we applied an under-sampling technique to balance the dataset and reassess the model's performance. The performance metrics for the under-sampled dataset are presented in Table \ref{tab:merged_metrics}. Following under-sampling, the model achieved a precision of $0.7971$, recall of $0.8293$, and F1-score of $0.8128$. This improvement suggests that the under-sampling technique enhanced the model's ability to classify exoplanets more effectively.

\begin{figure}
	\centering
	\includegraphics[width=1\columnwidth]{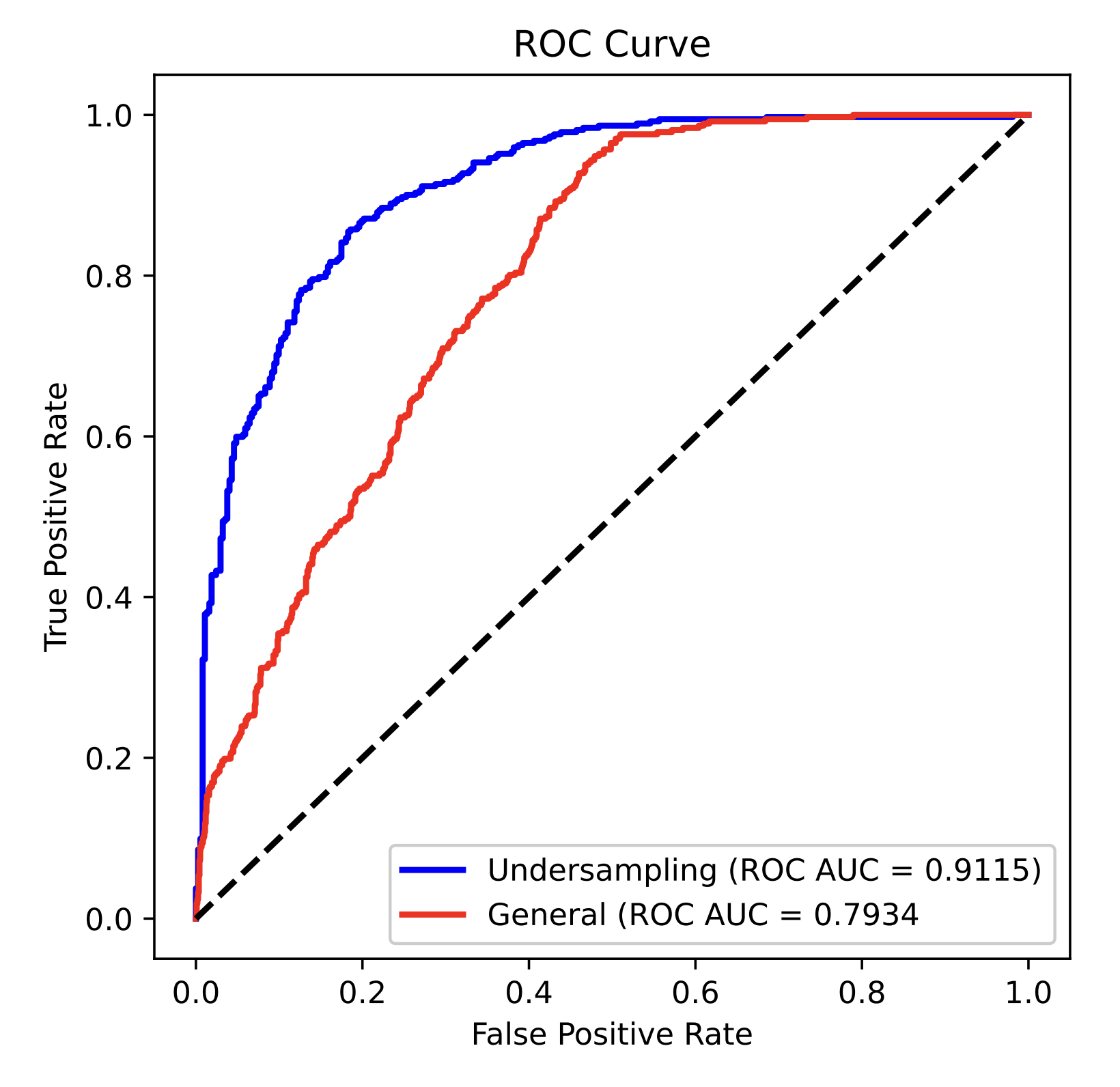}
    \caption{ROC Curve for the GAFs-based model, comparing the performance of the general model (red) and the under-sampled model (blue).}
   \label{fig:roc_curve_gaf}
\end{figure}

Similar to RPs, we also assessed the model's performance using the ROC curve. The ROC curve, shown in Figure \ref{fig:roc_curve_gaf}, provides a detailed illustration of the model's ability to distinguish between positive and negative classes. The model achieved a ROC value of $0.7934$, indicating strong predictive power. After applying the under-sampling technique, the ROC value improved to $0.9115$, highlighting a significant enhancement in the model's ability to discriminate between classes. This improvement suggests that the model benefits significantly from the under-sampling technique, enhancing its ability to classify exoplanets more effectively. \\
\begin{table}
	\caption{A comparative performance analysis of RPs and GAFs as input data following under-sampling.
	}
	\label{tab:merged_metrics}
	\centering
	\begin{tabular}{lcc}
		\hline\hline
		Metric & RPs & GAFs \\
		\hline
		Precision & 0.8426 & 0.7971 \\
		Recall & 0.8756 & 0.8293 \\
		F1 Score & 0.8587 & 0.8128 \\
		\hline
	\end{tabular}
\end{table}
\begin{table}[h!]
\centering
\caption{Comparative analysis of exoplanet classification studies.}
\label{state-of-the-art}
\begin{tabular}{lcc}
\hline\hline
\textbf{Type} & \textbf{Recall} & \textbf{Precision} \\ \hline
\cite{shallue2018identifying} & 0.95 & 0.93 \\ \hline
\cite{2022MNRAS.513.5505M} & 0.96 & 0.82 \\ \hline
Our method & 0.89 & 0.85 \\ \hline
\end{tabular}
\end{table}
We analyzed the results on balanced and imbalanced datasets to evaluate the model performance on both input representations. An under-sampling technique was applied to address the class imbalance and the performance was observed on each representation under varying data conditions. The results indicate that RPs demonstrate consistent performance, highlighting their effectiveness in managing class imbalance, whereas GAFs exhibit reasonable performance on balanced datasets, suggesting their suitability with evenly distributed classes but limited adaptability to imbalanced data.

We also compared the performance of our results with other state-of-the-art models based on deep learning, as shown in Table \ref{state-of-the-art}, and found that our model achieved comparable results (a recall of 0.89 with a precision of 0.85). Although \cite{shallue2018identifying} achieved higher recall and precision by using deep learning models with local light curves, our proposed model is better suited for learning long-term dependencies with respect to previous approaches. Similarly, \cite{2022MNRAS.513.5505M} proposed a planet detection method based on classical machine learning on the Kepler and TESS datasets; although the precision of their model is lower than that of the current study, the recall is higher, that indicates that without human supervision, the model is prone to make mistakes on unseen data. In contrast, our study has also provided a balanced trade-off between precision and recall that was shown by the variation in the mean of cross-validation, highlighting the effectiveness of ViT in exoplanet classification. Unlike previous studies, which primarily relied on deep network architectures, our approach used ViT, i.e., a deep network with an additional multi-head self-attention layer for effective feature extraction. We have also applied dropout regularization in the model to prevent overfitting by randomly dropping some of the output neurons. So, future research should explore more to fully use the potential of ViT in this field.

Our study highlights the use of RPs and GAFs as input representations using the ViT model, which has not been explored much in the field of exoplanet classification. The results demonstrate the efficacy of RPs in tasks related to classification by highlighting their ability to deliver reliable results even under class imbalance. Moreover, our investigation highlights the potential of RPs and ViT in capturing complex relationships within diverse data distribution compared to GAFs. 

Several limitations were identified in this study. While under-sampling addresses class imbalance with GAFs effectively, its impact on RPs was limited as the samples were randomly dropped, indicating the need for more advanced data balancing techniques. Additionally, RPs as input to the Vit model have not demonstrated significant improvements in performance as compared to the earlier studies shown in Table \ref{state-of-the-art}. However, their performance highlights the potential for further enhancement and modifications in the architecture of the ViT model. One possible approach to improve the model's performance is through a hybrid architecture of CNN and ViT, as by combining the strengths of both models, the overall performance of the model may be improved. Another key consideration is that using local light curves as the basic input for generating image representations may not fully exploit ViT's capabilities, so future exploration with global light curves as input could yield better results by providing a more comprehensive view of the data.
Furthermore, implementing a ViT model increases computational cost and model complexity. Therefore, future research should focus on optimizing and fine-tuning the model architecture to balance performance with computational efficiency.

	
	
	
	

\section{Summary and Conclusion}
\label{Conclusion}
In conclusion, we implemented a ViT model for exoplanet classification, focusing on a comparative analysis of image-based inputs. We evaluated the model's performance using two types of transformed inputs: RPs and GAFs. The results consistently revealed that images derived from RPs achieved higher classification results as compared to those obtained from GAFs. Additionally, the model that uses RPs is significant for handling imbalanced datasets effectively. This characteristic makes RPs a more suitable choice for scenarios involving data imbalance, in contrast to GAFs. For this application, RPs emerge as a more effective method for enhancing classification performance. Here we summarise the key points of our paper.
\begin{itemize}
    \item This study presents a new approach to classify exoplanets. It uses ViT to analyze image representations of Kepler light curve data. The data is transformed into RPs and GAFs, allowing the model to effectively capture the intricate temporal relationships within the data.
    \item The ViT model showed reasonable performance, especially when using RPs as input. It achieved an 89.46$\%$ recall rate and an 85.09$\%$ precision rate. These results emphasize the model's ability to accurately identify exoplanetary transits, demonstrating it as a potentially powerful tool for automated exoplanet detection.
    \item We found that RPs outperformed GAFs in classifying exoplanets. By using RPs, a higher performance in classification was achieved, indicating that they are more successful in capturing the essential characteristics required to differentiate exoplanets from non-exoplanetary occurrences.
    \item We also addressed the issue of class imbalance in the dataset using undersampling techniques and the model still performed well, but the smaller dataset size was seen as a limitation. 
    \item The evaluation metrics such as precision, recall, and F1-score provided a thorough assessment of the model's performance. The robustness of the model was also validated through ROC analysis. The RPs-based model achieved a high ROC value, indicating its reliability in accurately identifying true exoplanets.
    \item Although this study yielded promising results, it is important to acknowledge some limitations. Specifically, the training of ViT was computationally intensive and the improvements obtained through undersampling were modest. To address these issues, it is recommended that future research prioritize the development of more advanced techniques that can effectively capture temporal dependencies and manage class imbalance. Furthermore, optimizing the ViT architecture to reduce training time and computational resource requirements will enhance the scalability and applicability of the model in various astrophysical scenarios.
\end{itemize}


This research significantly contributes to the classification of exoplanets by introducing a significant approach using ViT with image-based time-series representations. The findings suggest that this methodology has the potential to improve the accuracy and reliability of exoplanet detection, which is vital for future space missions and automated data analysis in astronomy. Moving forward, it is important to develop more sophisticated techniques that can effectively capture the temporal complexities in RPs. Additionally, optimizing the ViT architecture to reduce computational demands and training time is crucial to expand the model's usability in different astrophysical contexts. By addressing these challenges, future research can build upon this work to create even more powerful and versatile tools for classifying exoplanets.


\section*{Acknowledgements}

In this work, we have used data from (a) the NASA Exoplanet Archive, run by the California Institute of Technology under an Exoplanet Exploration Program contract with NASA, and (b) the exoplanet.eu database which compiles and organizes data from various sources, including ground-based and space-based telescopes. We are thankful to the Inter-University Centre for Astronomy and Astrophysics (IUCAA) in Pune, India, for providing access of campus research resources under the IUCAA Visiting Associateship program. S.C. is funded by the European Union (ERC, UNVEIL, 101076613). Views and opinions expressed are however those of the author(s) only and do not necessarily reflect those of the European Union or the European Research Council. Neither the European Union nor the granting authority can be held responsible for them. S.C. acknowledges financial contribution from PRIN-MUR 2022YP5ACE.

\software{Numpy \citep{harris2020array}, Astropy \citep{robitaille2013astropy}, Scikit-learn \citep{pedregosa2011scikit}, Matplotlib \citep{hunter2007matplotlib}, Scipy \citep{virtanen2020scipy}, PyRQA \citep{rawald2017pyrqa}, Pandas \citep{2022zndo...3509134T}, Tensorflow \citep{2022zndo...7120930M}.}

\bibliography{biblio}{}
\bibliographystyle{aasjournal}
\appendix
\section{Model Architechture} \label{appendix:Model Architechture}
The model comprises $\sim$ 2.39 million parameters, and the configuration is as follows:
\begin{table}[h!]
\centering
\caption{Total number of trainable parameters in the ViT model.}
\begin{tabular}{|c|l|r|}
\hline\hline
\textbf{Sr. No.} & \textbf{Component} & \textbf{Parameters} \\
\hline\hline
1 & Patch Projection & 2,080 \\
\hline
2 & Positional Embedding & 18,432 \\
\hline
3 & Transformer Encoder Layers ($\times 10$) & \\
\hline
   & \quad Layer Normalization (before) & 64 \\
   & \quad Multi-Head Attention Layer & 16,800 \\
   & \quad Layer Normalization (after) & 64 \\
   & \quad Dense 1 & 2,112 \\
   & \quad Dense 2 & 2,080 \\
\hline
4 & Layer Normalization (final) & 64 \\
\hline
5 & MLP Head & \\
\hline
   & \quad Dense 1 & 67,584 \\
   & \quad Dense 2 & 2,098,176 \\
\hline
6 & Final Classification Layer & 2,050 \\
\hline\hline
   & \textbf{Total} & \textbf{2,399,586} \\
\hline
\end{tabular}
\end{table}

\end{document}